\documentclass[fleqn,twoside]{article}
\usepackage[headings]{espcrc2}

\readRCS
$Id: espcrc2.tex,v 1.2 2004/02/24 11:22:11 spepping Exp $
\usepackage{graphicx}
\usepackage[figuresright]{rotating}

\newcommand{\AmS}{{\protect\the\textfont2
  A\kern-.1667em\lower.5ex\hbox{M}\kern-.125emS}}

\newcommand{\BlackHat}{{\sc BlackHat}}
\newcommand{\SHERPA}{{\sc SHERPA}}
\newcommand{\AMEGIC}{{\sc AMEGIC++\/}}
\newcommand{\SISCone}{{\sc SISCone}}

\newcommand{\pt}{$p_T$}

\def\Wjjj{$W\,\!+\,3$}
\def\Wjjjj{$W\,\!+\,4$}
\def\Wjx{$W\,\!+\,1,2$}

\def\Zjjj{$Z\,\!+\,3$}

\def\Zgamjjj{$Z,\gamma^*\,\!+\,3$}

\def\Zgamjjja{$Z,\gamma^*\,\!+\,1,2,3$}

\def\Zjnp1{$Z\,\!+\,(n+1)$}

\def\Zgamjnp1{$Z,\gamma^*\,\!+\,(n+1)$}
\def\Zjnm1{$Z\,\!+\,(n-1)$}
\def\Zgamjnm1{$Z,\gamma^*\,\!+\,(n-1)$}

\def\Vjjj{$V\,\!+\,3$}
\def\Vjjjj{$V\,\!+\,4$}

\def\jet{{\rm jet}}
\def\eps{\epsilon}

\newif\ifdraft
\drafttrue
\newif\ifpreprint
\preprinttrue

\def\fig#1{fig.~{\ref{#1}}}
\def\Fig#1{Fig.~{\ref{#1}}}

\def\eq<ns#1#2{eqs.~(\ref{#1}) and~(\ref{#2})}

\title{\vspace{-5cm} \rule{1cm}{0cm}
\noindent {\small\rm SLAC--PUB--14105  \hskip 1.62cm
UCLA/10/TEP/104 \hskip 1.62cm
MIT-CTP 4150  \hskip 1.62cm
SB/F/380-10\\
\hskip 1.2 cm 
IPPP/10/34 \hskip 1.13cm \null
Saclay IPhT--T10/065 \hskip 1.13cm 
NIKHEF-2010-013 \hskip 1.13cm
CERN-PH-TH/2010-109
}  \rule{0cm}{0cm}
 \vspace{2cm} \\
Vector Boson + Jets with BlackHat and Sherpa}

\author{C.~F.~Berger\address{Center for Theoretical
Physics, Massachusetts Institute of Technology,
   Cambridge, MA 02139, USA},
Z.~Bern\address[UCLA]{Department of Physics and Astronomy, UCLA, 
   Los Angeles, CA 90095-1547, USA},
L.~J.~Dixon\address[SLAC]{SLAC National Accelerator Laboratory, 
   Stanford University, Stanford, CA 94309, USA},
F.~Febres Cordero\address[Caracas]{Universidad Sim\'on Bol\'{\i}var, Departamento de
F\'{\i}sica, Caracas 1080A, Venezuela},
D.~Forde\address{Theory Division, Physics Department, CERN, CH--1211 Geneva 23, 
    Switzerland}\address{NIKHEF Theory Group, Science Park 105, NL--1098~XG
  Amsterdam, The Netherlands},
T.~Gleisberg\addressmark[SLAC],
H.~Ita\addressmark[UCLA],
D.~A.~Kosower\address{Institut de Physique Th\'eorique, CEA--Saclay,
          F--91191 Gif-sur-Yvette cedex, France},
D.~Ma\^{\i}tre\address[IPPP]{Department of Physics, University of Durham,
          DH1 3LE, UK}\thanks{Presenter at {\it Loops and Legs in 
Quantum Field Theory 2010}, W\"{o}rlitz, Germany, April 25-30, 2010 }}
       
\runtitle{Vector Boson + Jets with BlackHat and Sherpa}
\runauthor{D. Ma\^{\i}tre et al.{}}

\begin{document}


\begin{abstract}
We review recent NLO QCD results for $W,Z + 3$-jet production at
hadron colliders, computed using \BlackHat{} and \SHERPA.
We also include some new results for \Zjjj-jet production at the
LHC at 7 TeV. We report new progress towards the NLO cross section
for \Wjjjj-jet production.  In particular, we show that the virtual
matrix elements produced by \BlackHat{} are numerically stable.
We also show that with an improved integrator and tree-level matrix 
elements from \BlackHat{}, \SHERPA{} produces well-behaved real-emission
contributions.  As an illustration, we present the real-emission
contributions---including dipole-subtraction terms---to the \pt{}
distribution of the fourth jet, for a single subprocess with the
maximum number of gluons.
\end{abstract}

\maketitle

\section{Introduction}

In the coming years a major theoretical task will be to provide
reliable predictions for hard-scattering processes at the LHC.  The
start of the LHC era in particle physics opens new opportunities to
confront data with theoretical predictions at scales well beyond those
probed in previous colliders.  The first observation at the LHC of
vector-boson production marks an important milestone.  
Processes involving vector bosons in association with jets are central
to the physics program of the LHC.  They are backgrounds to Higgs and top
physics, as well as to many signals of new physics.  Theoretical
predictions can play an important role in experimentally-driven
determinations of backgrounds, improving extrapolations to signal
regions.  $Z$ and $W$ production can play complementary roles in such
determinations of backgrounds.  At low luminosity, $W$ production can
be used to calibrate estimates of $Z$ production, because its
cross section is higher; while at high luminosity, $Z$ production can
be used to calibrate $W$ production because of the cleanliness of
measuring lepton pairs.  The good theoretical understanding of
vector-boson production also allows its use to measure the luminosity.

Next-to-leading-order (NLO) QCD is a key tool for confronting theory
with experiment.  This order in perturbation theory is the first to
provide quantitatively reliable results, as it is the first order at
which quantum corrections compensate the renormalization- and
factorization-scale dependence in the strong coupling $\alpha_s$.
That scale dependence in leading-order (LO) predictions increases with
the number of jets because of the increasing number of powers of
$\alpha_s$.  At LO the scale dependence can be quite large, on the
order of 50\% for three or more jets.  Furthermore, LO results
may not model the shapes of distributions correctly.  Although NLO
computations are more challenging, in general they yield results with
better reliability and agreement with measurements (see {\it e.g.}
refs.~\cite{LesHouches,WCDF,ZCDF,PRLW3BH,W3jDistributions,%
EMZW3j,ZD0,TEVZ,HeavyQuark6Pt}).

For many years the bottleneck blocking NLO computations of
vector-boson production with three or more jets has been the
difficulty of evaluating the required one-loop amplitudes.  The
unitarity method~\cite{UnitarityMethod,OnShellReviews},
along with various important
developments~\cite{GeneralizedUnitarity,LoopRecursion}, has broken
past this barrier, producing NLO results for vector-boson
production in association with three
jets~\cite{W3jDistributions,TEVZ}.  (See
refs.~\cite{PRLW3BH,EMZW3j} for other unitarity-based results,
using various leading-color approximations to \Wjjj{} jets at NLO.)
NLO computations of six-point processes involving heavy quarks have
also been carried out recently, using both Feynman diagrams and on-shell
methods~\cite{HeavyQuark6Pt}.

In this talk we briefly review the NLO results for $W,Z+3$-jet
production presented in refs.~\cite{W3jDistributions,TEVZ}.  We also
present new results on $Z$ boson production at the LHC, all at
the current center-of-mass energy of 7~TeV.
Finally, we
demonstrate the feasibility of computing \Wjjjj-jet production at NLO. 
The latter process is key for new physics searches at the LHC, and
from a theoretical point of view goes one jet beyond the
2007 Les Houches ``experimenters' wishlist''~\cite{LesHouches}.
We show that the virtual matrix elements
computed by the \BlackHat{} library are numerically stable for the
most complicated partonic subprocesses, those involving the maximum number of
gluons.  We also show that, after augmenting \SHERPA{} with tree-level
matrix elements from \BlackHat{} and incorporating an updated
integrator~\cite{ImprovedIntegrator}, we can obtain smooth
distributions for the real-emission terms.  As an illustration
we present the infrared-subtracted real-emission contribution for a
single subprocess.

\section{Calculational Setup}

NLO corrections to cross sections require the evaluation of virtual
and real-emission contributions.  We evaluate the virtual amplitudes
using the \BlackHat{} library~\cite{BlackHatI,W3jDistributions},
a numerical implementation of on-shell methods~\cite{UnitarityMethod,%
OnShellReviews,GeneralizedUnitarity,LoopRecursion} at one loop.
For the real emission contributions we use
the Catani--Seymour dipole subtraction method~\cite{CS}, as
implemented~\cite{AutomatedAmegic} in the
program~\AMEGIC{}~\cite{Amegic} (part of the \SHERPA{}
framework~\cite{Sherpa}) to cancel the infrared divergences arising in
the phase-space integration.  For subprocesses with up to eight
external legs we use \AMEGIC{} to generate any required tree
amplitudes.  We also use the \SHERPA{} package to carry out the
numerical integration over phase space.  Details of the setup can be
found in refs.~\cite{BlackHatI,W3jDistributions,TEVZ}.


\section{NLO Results for Vector-Boson Production with 
Three Jets}

\begin{figure}[tbh]
\begin{center}
\includegraphics[clip,scale=0.37]{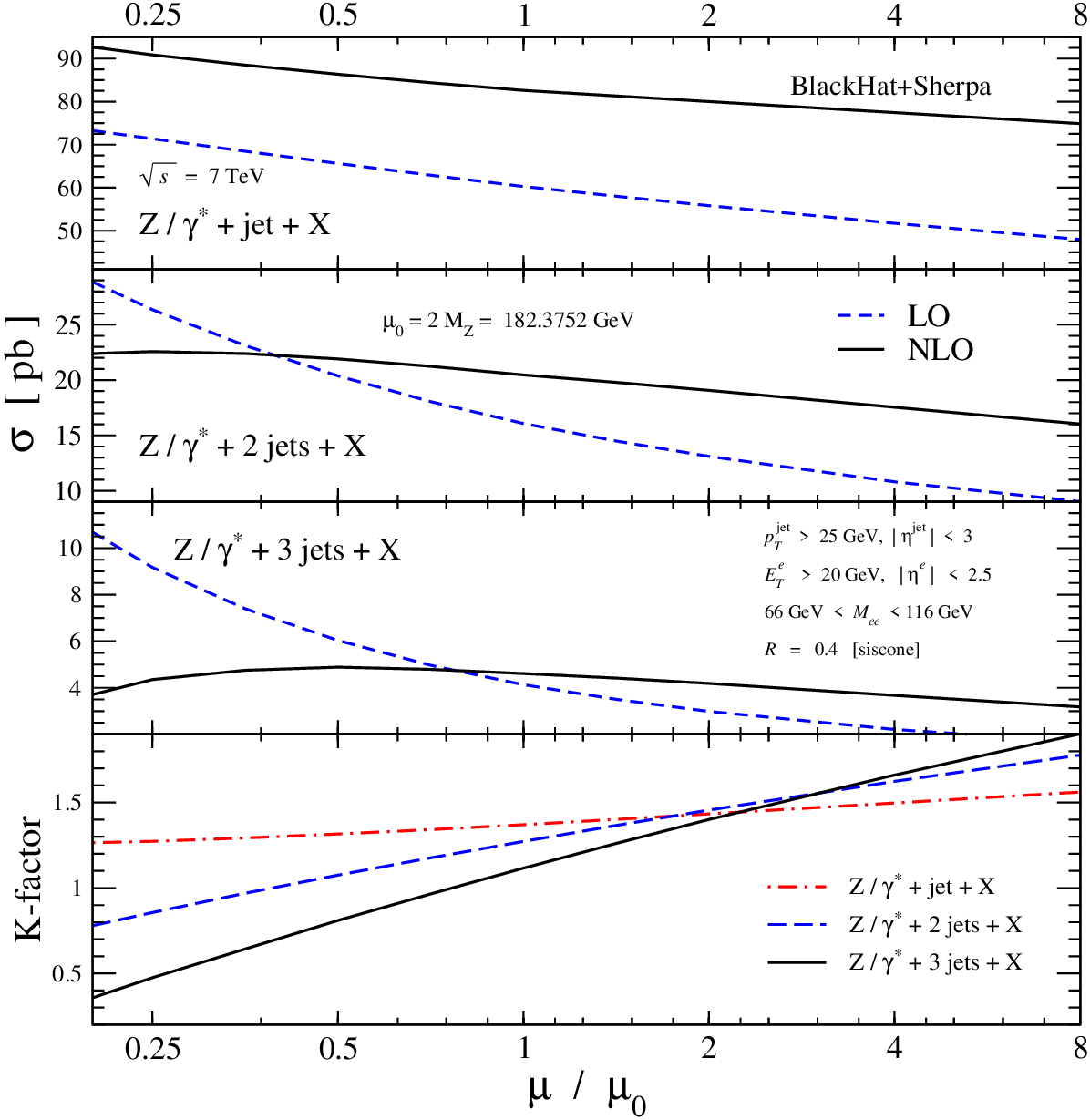}
\end{center}

\vskip -1. cm
\caption{  \small
The scale dependence of the LO (dashed blue) and NLO (solid
  black) cross sections for \Zgamjjja-jet production at the LHC running
  at 7 TeV,
  as a function of the common renormalization and factorization scale
  $\mu$, with $\mu_0 = M_Z$.  The bottom panel shows the $K$ factors, or ratio
  between the NLO and LO result, for each of the three cases: 1 jet
  (dot-dashed red), 2 jets (dashed blue), and 3 jets (solid black).}
\label{ZJetsScaleVariationSISConeFigure}
\end{figure}

In ref.~\cite{W3jDistributions} we presented a detailed study of
\Wjjj-jet production at NLO, at the Tevatron and LHC.  This study
exhibited the expected scale-dependence reduction, and correspondingly
better theoretical precision, compared with LO predictions.
In addition, we found that at the LHC, the
$W^+$ and $W^-$ are {\it both} preferentially polarized left-handed at
large transverse momentum.  This effect is also present in \Wjx-jet 
production, and at LO as well (see also ref.~\cite{BHRadcor09}).
Such polarization will be largely, if
not completely, absent in $W$ bosons emerging from decays of top quarks
or other new heavy states.  Accordingly, it should be useful in
setting experimental cuts that will distinguish between such daughter
$W$s and ``prompt'' $W$s emitted directly in the short-distance process.
In this Proceeding, we will discuss the more recent NLO
computation of \Zjjj-jet production at the 
Tevatron~\cite{TEVZ}, for which data already exists~\cite{ZCDF,ZD0},
as well as new NLO predictions for the current LHC run.

The dependence of NLO cross sections on the common renormalization 
and factorization scale, $\mu = \mu_R = \mu_F$, 
is greatly reduced with respect to LO,
as illustrated in
\fig{ZJetsScaleVariationSISConeFigure} for \Zgamjjja-jet
production at the LHC at 7 TeV.  For these plots the \SISCone{} jet
algorithm~\cite{SISCONE} is used with cone size $R=0.4$ and merging parameter
$f=0.75$; the lepton and jet cuts are indicated on the figure.  (The
case of three jets uses a leading-color approximation with the
same separation of leading and subleading terms described in 
ref.~\cite{TEVZ}; we expect it to be accurate to
within a few percent.)  Indeed, the scale-dependence reduction becomes
more pronounced as the number of jets increases, in accord with the
increasing variation at LO due to increasing powers of $\alpha_s(\mu)$.
The corresponding plot for production at the Tevatron shows
similar behavior~\cite{TEVZ}.

\begin{figure*}[tbh]
\begin{center}
\begin{minipage}[b]{1.\linewidth}
\hskip .8 cm 
\includegraphics[clip,scale=0.33]{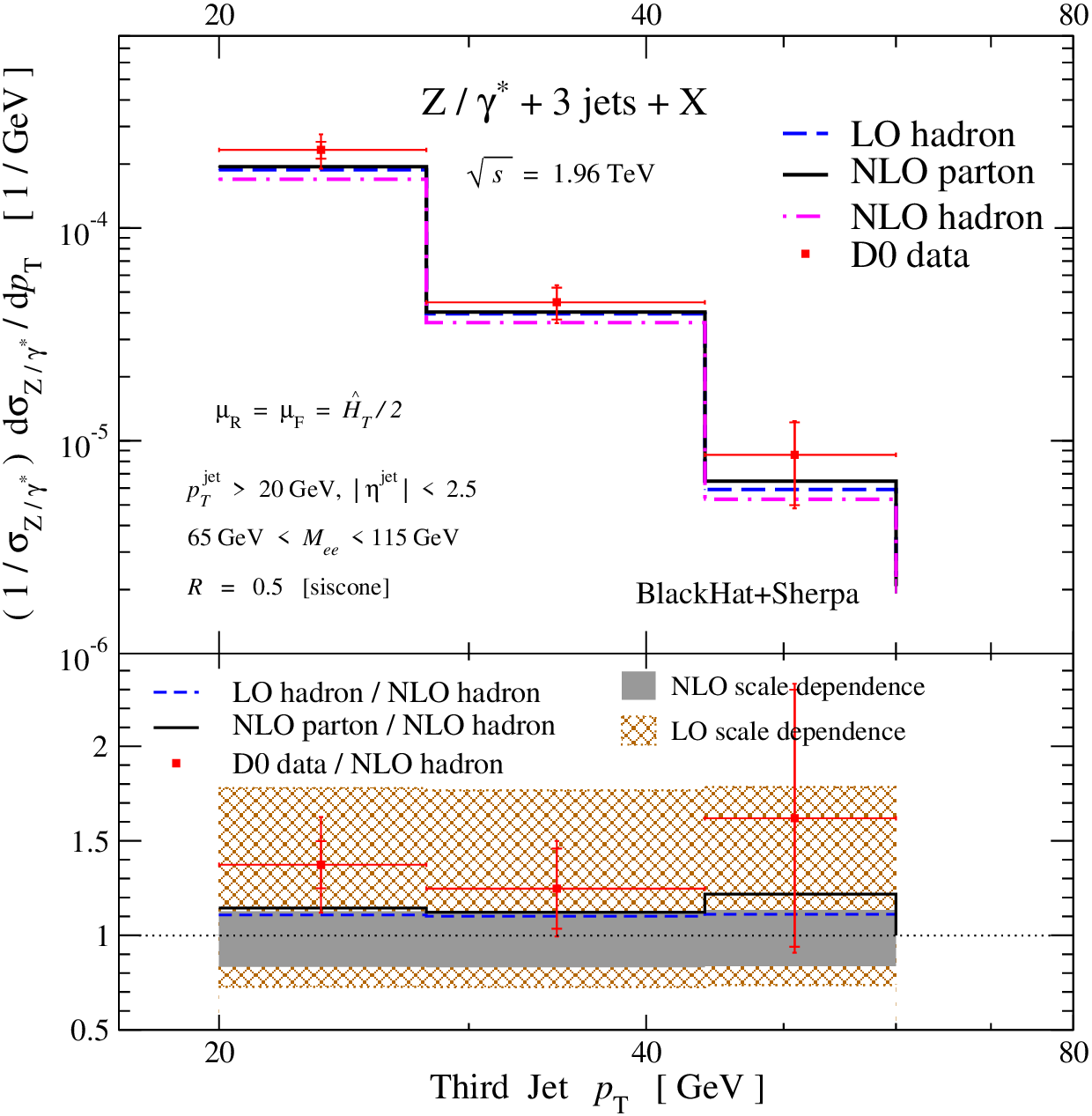}
\hskip 1 cm 
\includegraphics[clip,scale=0.33]{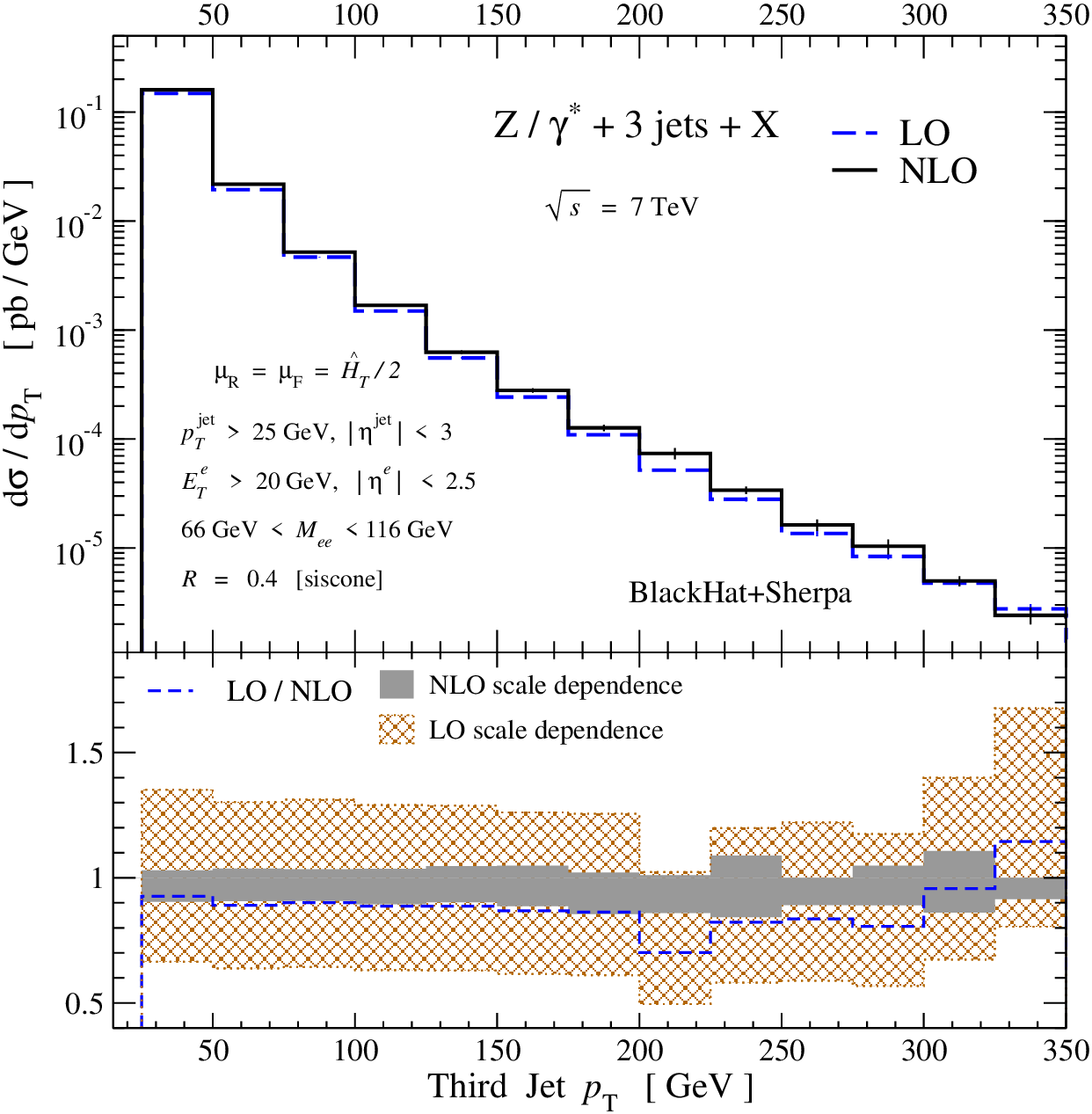} 
\end{minipage}
\end{center}

\vskip -1 cm 
\caption{\small
  The left plot shows $1/\sigma_{Z,\gamma^*} \times d \sigma/d p_T$ 
  for the third-hardest jet in \Zgamjjj-jet production,
  comparing D0 data against LO and NLO predictions.  In the upper
  panel the parton-level NLO distributions are the solid (black)
  histograms, while the NLO distributions corrected to hadron level
  are given by dash-dot (magenta) histograms.  The D0 data is
  indicated by the (red) points. The LO predictions corrected to
  hadron level are shown as dashed (blue) lines.  The lower panel
  shows the distribution normalized to the full hadron-level NLO
  prediction. The scale-dependence bands in the lower panels are
  shaded (gray) for NLO and cross-hatched (brown) for LO.  The right
  plot shows the \pt{} distribution for the third
  jet at the LHC, with the indicated cuts.  
  The complete color dependence is included in this plot, but no corrections 
  to hadron level are applied.}

\label{Figure-Z23jet-D0data-siscone-HT}
\end{figure*}

In the left panel of \fig{Figure-Z23jet-D0data-siscone-HT} we compare
the distribution in \pt{} for the third-hardest jet (\pt-ordered)
in \Zgamjjj-jet production, computed at LO and NLO~\cite{TEVZ}, 
to D0 data.  D0 provided results for two data
selections~\cite{ZD0}, with the $Z$ decaying into and
electron-positron pair.  \Fig{Figure-Z23jet-D0data-siscone-HT} uses
the primary data selection based on the cuts,
$p_T^\jet > 20~{\rm GeV}, |\eta^\jet| < 2.5,$ and
$65 {\ \rm GeV} < M_{e e} < 115 {\ \rm GeV}$.
D0 defined jets using the D0 Run II midpoint jet
algorithm~\cite{D0jetAlgorithm}, with a cone size of $R=0.5$ and a
merging/splitting fraction of $f=0.5$. Because this algorithm is not
infrared-safe, we use instead the \SISCone{} algorithm~\cite{SISCONE},
with the same parameters, and a scale choice $\mu=\hat{H}_T/2$.  We
also included non-perturbative corrections estimated by D0~\cite{ZD0}. 

The figure demonstrates good agreement between NLO theory and
experiment, despite a number of issues, including the
difficulty of obtaining precise nonperturbative corrections,
and differences in the jet algorithms.  Further details of comparisons to
CDF and DO data for $Z$-boson production in association with jets may
be found in ref.~\cite{TEVZ}. 

The right panel of \fig{Figure-Z23jet-D0data-siscone-HT} shows the
same distribution at the LHC (except that it is
not normalized to $\sigma_{Z,\gamma^*}$).  The experimental cuts are
labeled on the plot. We again use the \SISCone{} jet algorithm,
which (along with anti-$k_T$) is one of the infrared-safe
jet algorithms adopted by the LHC experiments.  Our setup allows
results for different algorithms and cone sizes to be computed at 
the same time.  It should be possible to minimize non-perturbative
effects by choosing appropriate cone sizes~\cite{Jetography},
perhaps in a $p_T$-dependent fashion.


\section{Towards $W + 4$ Jets at NLO}

\begin{figure}[tbh]
\begin{center}
\includegraphics[clip,scale=0.55]{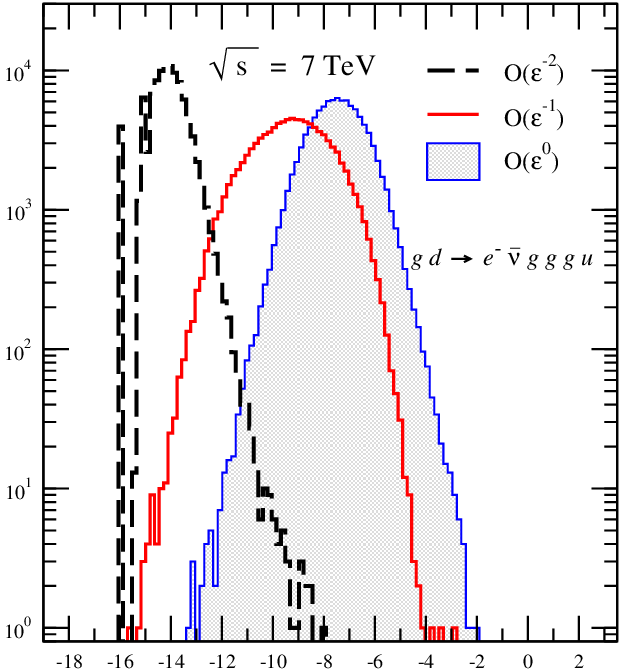}
\end{center}

\vskip -.8 cm 
\caption{\small The distribution of the relative error in the leading-color
  virtual cross section for the subprocesses $g d \rightarrow e^- \bar
  \nu g g g u $.  The horizontal axis is the logarithm of the relative
  error between an evaluation by \BlackHat{}, running in production
  mode, and a higher precision target. The vertical axis shows the number
  of phase-space points out of 100,000 that have the corresponding
  error.  The dashed (black) line shows the $1/\eps^2$ term; the solid
  (red) curve, the $1/\eps$ term; and the shaded (blue) curve, the
  finite ($\eps^0$) term.}
\label{W4StabilityFigure}
\end{figure}

\Wjjjj-jet production has been an important background since the early
days of the Tevatron.  There, it was the dominant background to
$t{\bar t}$ production; at the LHC, it will be an important background
to many new-physics searches, and will continue to be important to
precision top-quark measurements.  Its computation beyond LO
represents an important challenge to theorists.  We report here on
progress towards an NLO calculation of this process.  

A crucial issue in the approach pursued in \BlackHat{} is numerical
stability of the virtual contributions.  In \fig{W4StabilityFigure},
we illustrate the stability of the leading-color virtual corrections
to the squared matrix element, $d\sigma_V$, summed over colors and
over all helicity configurations for the subprocess $g d \rightarrow
e^- \bar \nu gg g u $.  The horizontal axis of \fig{W4StabilityFigure}
shows the logarithmic error,
$$
\log_{10}\left({|d \sigma_V^{\rm BH}- d \sigma_V^{\rm target}|} /
           | d \sigma_V^{\rm target}| \right), \nonumber
$$
for each of the three components: $1/\epsilon^2$, $1/\epsilon$ and
$\epsilon^0$, where $\eps = (4-D)/2$ is the dimensional regularization
parameter. In this expression $\sigma_V^{\rm BH}$ is the cross section
computed by \BlackHat{} as it normally operates for production runs,
whereas $\sigma_V^{\rm target}$ is a target value computed by
\BlackHat{} using higher-precision arithmetic of at least 32 digits.
The phase-space points
are selected in the same way as those used to compute cross sections.
As seen in \fig{W4StabilityFigure}, no point has a relative error
larger than about 1\%; this level of
precision is more than adequate for phenomenology.


\begin{figure}[htb]
\begin{center}
\includegraphics[scale=0.35]{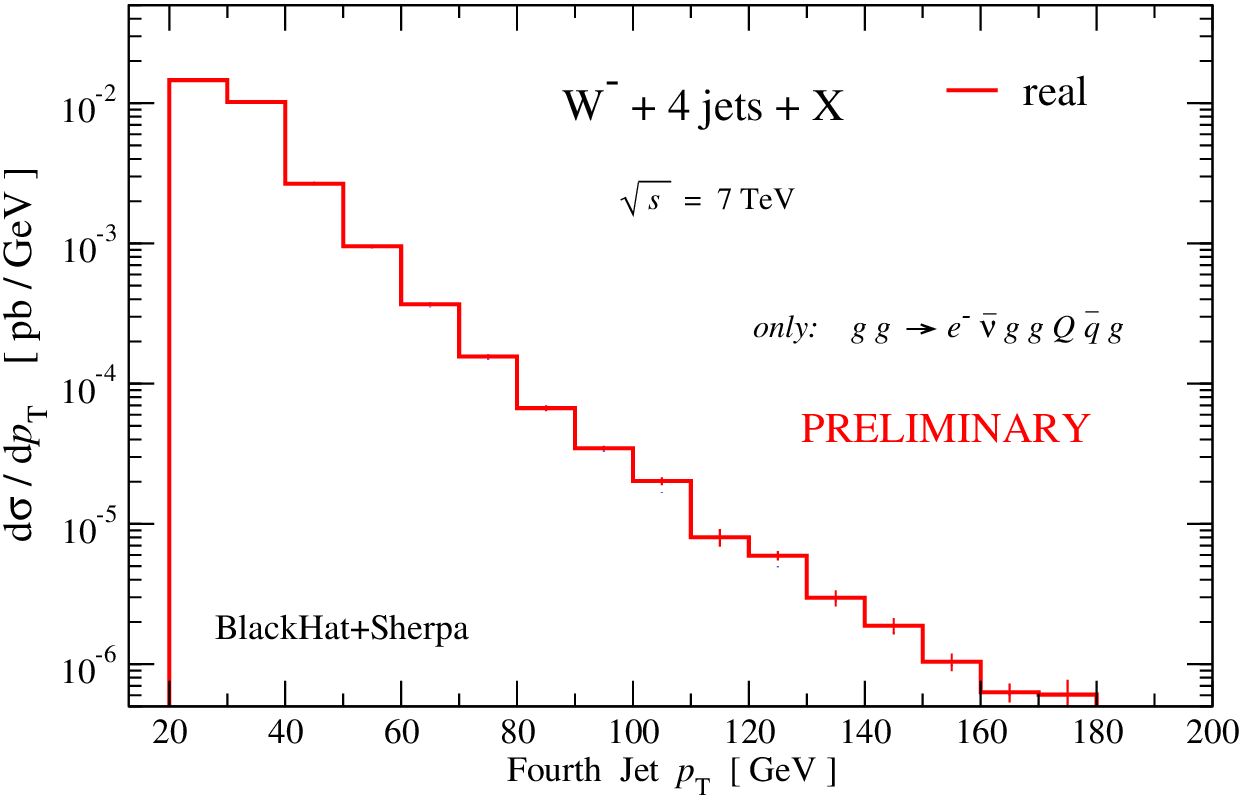}
\end{center}

\vskip -1 cm 

\caption{The real part of the NLO $p_T$ distribution for the fourth
  jet in \Wjjjj-jet production at the LHC. Only the subprocess $gg
  \rightarrow e^- \bar \nu g g Q \bar q g$ is included here.  The thin
  vertical lines, where visible, indicate numerical integration
  uncertainties.}
\label{Wm4RealPtFigure}
\end{figure}

What about the real corrections?  In \fig{Wm4RealPtFigure} we show 
the real-emission contributions, including dipole subtractions,
for the distribution in the fourth-jet \pt.  Although these 
contributions alone are not physically
meaningful (they need to be combined with the 
analytically-integrated dipole terms and the virtual corrections,
and at the very least summed over other single-quark-line subprocesses),
they do serve to illustrate that our
\SHERPA-based integration setup has them under control.  Here 
we used the anti-$k_T$ algorithm for $R=0.4$,
with a jet cut of $p_T > 25$ GeV,
along with other cuts on the leptons and the jet rapidities.
For this plot, the real terms were sampled at $2\times 10^8$
phase-space points, yielding the rather small integration errors
indicated by the thin vertical lines.  Two key improvements in our
setup, which make the real contributions feasible, are an improved
integrator~\cite{ImprovedIntegrator,TEVZ} and more efficient
nine-point tree-level amplitudes generated by \BlackHat{} using
on-shell recursion relations~\cite{BCFW}.

\section{Conclusion}

A publicly available version of \BlackHat{} is under construction and is
being tested in a variety of projects (see {\it
e.g.}~ref.~\cite{RikkertRadcor}).  This version uses the Binoth Les
Houches interface~\cite{BinothInterface}. It has been tested with both
C++ and Fortran clients.  We intend the public version to provide all
processes that have been thoroughly vetted using the full \BlackHat{}
code.  We expect significant further gains in efficiency after
implementing a variety of additional optimizations, such as taking
advantage of phase-space symmetries to reduce the amount of
computation.

In summary, \Vjjj-jet production (where $V$ is a
vector boson) at hadron colliders is under excellent control
at NLO, using \BlackHat{} and \SHERPA{}, allowing
comparisons~\cite{PRLW3BH,W3jDistributions,TEVZ} to Tevatron
data~\cite{WCDF,ZCDF,ZD0} and predictions for the LHC.
The frontier has now shifted to \Vjjjj-jet production.  In this
Proceeding, we presented significant new progress towards this goal, and
demonstrated for a selected subprocess that the programs are able to
execute these complex calculations. After further optimization and
testing we plan to compute \Vjjjj-jet production at NLO
at the LHC.  More generally, we look forward to applying these tools
to a wide range of studies of LHC physics.

\section*{Acknowledgments}

This research was supported by the US Department of Energy under
contracts DE--FG03--91ER40662, DE--AC02--76SF00515 and
DE--FC02--94ER40818.  DAK's research is supported by the European
Research Council under Advanced Investigator Grant ERC--AdG--228301.
HI's work is supported by a grant from the US
LHC Theory Initiative through NSF contract PHY-0705682.
This research used resources of Academic Technology Services at UCLA,
PhenoGrid using the GridPP infrastructure, and the National Energy
Research Scientific Computing Center, which is supported by the Office
of Science of the U.S. Department of Energy under Contract
No. DE--AC02--05CH11231.

\end{document}